\documentclass[superscriptaddress,preprint]{revtex4}
\usepackage{mathrsfs}
\usepackage{amssymb}
\usepackage[tbtags]{amsmath}
\usepackage{graphicx}
\usepackage{epsfig,graphicx,times}
\usepackage{color}

\begin{document}

\title{Jaynes-Cummings Models with trapped electrons on liquid Helium}
\author{Miao Zhang, H.Y. Jia and L.F. Wei\footnote{weilianfu@gmail.com}}
\affiliation{Quantum Optoelectronics Laboratory and Institute of
Modern Physics, Southwest Jiaotong University, Chengdu 610031,
China}
\date{\today}

\begin{abstract}
Jaynes-Cummings model is a typical model in quantum optics and has
been realized with various physical systems (e.g, cavity QED,
trapped ions, and circuit QED etc..) of two-level atoms interacting
with quantized bosonic fields. Here, we propose a new implementation
of this model by using a single classical laser beam to drive an
electron floating on liquid Helium.
Two lowest levels of the {\it vertical} motion of the electron acts
as a two-level ``atom", and the quantized vibration of the electron
along one of the {\it parallel} directions, e.g., $x$-direction,
serves the bosonic mode.
These two degrees of freedom of the trapped electron can be coupled
together by using a classical laser field. If the frequencies of the
applied laser fields are properly set, the desirable Jaynes-Cummings
models could be effectively realized.

PACS number(s): 42.50.Dv, 42.50.Ct, 73.20.-r.
\end{abstract}

\maketitle

Jaynes-Cummings model (JCM), describing the basic interaction of a
two-level atom and a quantized electromagnetic field, is a
cornerstone for the treatment of the interaction between light and
matter in {\it Quantum Optics}~\cite{JCM0}. This model can explain
many quantum phenomena, such as the collapses and revivals of the
atomic population inversions, squeezing of the quantized field, and
the atom-cavity entanglement.
Furthermore, recent experiments show that the JCMs can be implicated
in quantum-state engineering and quantum information processing,
e.g., generation of Fock states~\cite{Fock} and entangled
states~\cite{entangled}, and the implementations of quantum logic
gates~\cite{gates}, etc.. Originally, JCM is physically implemented
with a cavity quantum electrodynamics (QED) system (see,
e.g.,~\cite{QED}). Certainly, there has been also interest to
realize the Jaynes-Cummings Hamiltonian with other physical systems.
A typical system is a cold ion trapped in a Paul trap and driven by
classical laser beams~\cite{trapped-ion,Rev.Mod.trapped.ion,JCMs1}.
There, the interaction between two selected internal electronic
levels and the external vibrational mode of the ion can be induced.
Under the so-called Lamb-Dicke (LD) limit and the well-known
rotating-wave approximation, the desirable JCM (or anti-JCM) can be
realized by setting the applied laser frequencies with the suitable
red (or blue) sideband excitations.

Recently, Platzman and Dykman have proposed that the electrons
floating on liquid Helium could be utilized to implement quantum
computation~\cite{Science,PRB}. In this proposal, electrons are
trapped on the surface of liquid Helium and controlled by a series
of external electric fields, which are generated by the
micro-electrodes set below the liquid Helium.
These electrons are effectively coupled together via their Coulomb
interactions. By applying microwave radiation to these electrons
from the micro-electrodes, their quantum states could be coherently
controlled.
Due to its scalability, easy manipulation, and relative long
coherence time, this system has been paid much attention in recent
years for quantum information processing (see,
e.g.,~\cite{Science,PRB,electron1,APL,electron2}).

In this paper, we further show, theoretically, that {\it an electron
floating on liquid Helium} could also be utilized to realize the
desirable JCMs. Inspired by the idea of implementing JCMs with
trapped ions, we use a classical laser field to couple the vertical
and parallel motional degrees of freedom of the electron on liquid
Helium (similar to the laser-assisted coupling between the internal
and external states of trapped ions).

We consider an electron floating on the surface of liquid Helium
(e.g., $^{4}\text{He}$). The electron is weakly attracted by the
dielectric image potential and strongly repulsed by the Pauli
potential (i.e., Pauli exclusion principle), with about $1$eV
potential barrier, to prevent it from penetrating into the liquid
Helium. As a consequence, the electron's motion normal to the liquid
Helium surface can be approximately described by a one-dimensional
($1$D) hydrogen with the following potential~\cite{Rev.Mod.Phys.}
\begin{equation}
V(z)=\left\{
\begin{array}{cc}
-\frac{\Lambda e^2}{z} \,\,\,\,\,\,\,z>0, \\
\\
+\infty \,\,\,\,\,\,\,\,\,\,z\leq0.
\end{array}
\right.
\end{equation}
Where, $e$ is electron (with mass $m_e$) charge, $z$ is the distance
above liquid Helium surface, and
$\Lambda=(\varepsilon-1)/4(\varepsilon+1)$ with $\varepsilon=1.0568$
being dielectric constant of liquid $^{4}\text{He}$.
The energy levels associated with this motion form a hydrogen-like
spectrums $E_n=-\Lambda^2e^4m_e/2n^2\hbar^2\approx-0.00065/n^2$ eV,
which has been experimentally observed~\cite{Grimes}, and the
corresponding wave functions can be written as~\cite{eigen-states}
\begin{equation}
\psi_n(z)=2n^{-\frac{5}{2}}r_B^{-\frac{3}{2}}
z\exp[-\frac{z}{nr_B}]L_{n-1}^{(1)}(\frac{2z}{nr_B}),
\end{equation}
with the Bohr radius $r_B=\hbar^2/(m_ee^2\Lambda)\approx76${\AA} and
Laguerre polynomials
\begin{equation}
L_n^{(\alpha)}(x)=\frac{e^xx^{-\alpha}}{n!}\frac{d^n}{dx^n}[e^{-x}x^{n+\alpha}].
\end{equation}
\begin{figure}[tbp]
\includegraphics[width=9.5cm]{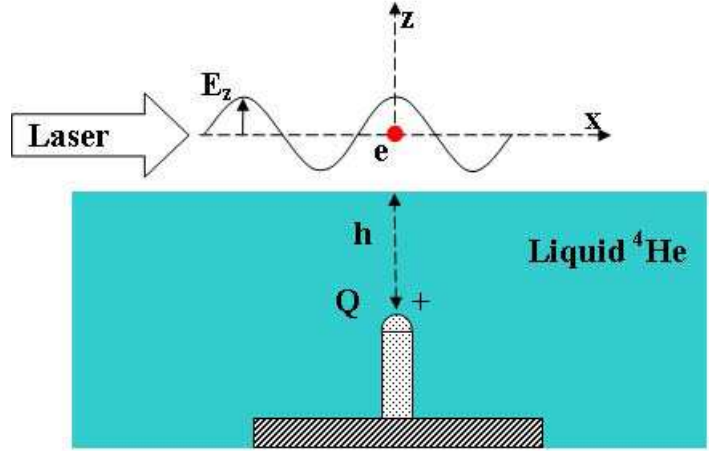}
\caption{(Color online) A sketch of an electron confined by a
micro-electode $Q$ submerged by the depth $h$ beneath the Helium
surface and driven by a classical laser field propagating along the
$x$-direction.}
\end{figure}
Beside the image potential (1), the electron is also trapped by
another potential generated by the charge $Q$ on the
micro-electrode, which is located at $h$ beneath the liquid Helium
surface~\cite{PRB}. The configuration of our model is shown in
Fig.~1.
For simplicity, on the Helium surface the electron is assumed to be
effectively constrained to move only along the $x$-axes.
Therefore, under the usual condition: $z, x<< h$, the total
potential of the electron can be effectively approximated
as~\cite{PRB}
\begin{equation}
U(z,x)\approx-\frac{\Lambda e^2}{z}+eE_\bot z+\frac{1}{2}m_e\nu^2x^2
\end{equation}
with $E_\perp\approx Q/h^2$ and
$\nu\approx\sqrt{eQ/(m_eh^3)}=\sqrt{eE_\perp/(m_eh)}$\,. This
indicates that the motions of the trapped electron can be regarded
as a 1D {\it Stark-shifted} hydrogen along the $z$-direction, and a
harmonic oscillation along the $x$-direction.
Following Dykman et.al.~\cite{PRB}, only two lowest levels (i.e.,
the ground state $|g\rangle$ and first excited state $|e\rangle$) of
the 1D Stark-shifted hydrogen are considered. As a consequence, the
Hamiltonian describing these two uncoupled degrees of freedom of the
electron reads
\begin{equation}
\hat{H}_0=\hbar\nu(\hat{a}^\dagger\hat{a}+\frac{1}{2})+\frac{\hbar
\omega_0}{2}\hat{\sigma}_z.
\end{equation}
Here, $\hat{a}^\dagger$ and $\hat{a}$ are the bosonic creation and
annihilation operators of the vibrational quanta (with frequency
$\nu$) of the electron's oscillation along the $x$-direction.
$\hat{\sigma}_z=|e\rangle\langle e|-|g\rangle\langle g|$ is the
Pauli operator. The transition frequency $\omega_0$ is defined by
$\omega_0=(E_e-E_g)/\hbar$ with $E_g$ and $E_e$ being the
corresponding energies of the lowest two levels, respectively.

In order to couple the above two uncoupled degrees of freedom of the
electron, we now apply a classical laser beam $\mathscr{E}(x,t)$,
propagating along the $x$-direction, to the trapped electron (see
Fig.~1). This is similar to the approach in ion trap system for
coupling the external and internal degrees of freedom of the
ion~\cite{Rev.Mod.trapped.ion}.
Suppose that the applied laser beam (of wave-vector $k_l$, amplitude
$E_z$, frequency $\omega_l$ and initial phase $\phi_l$) takes the
form $\mathscr{E}(x,t)=E_z\hat{z}\cos(k_lx-\omega_lt+\phi_l)$, i.e.,
its electric field is $z$-direction polarization, then the
Hamiltonian of the driven electron floating on the Helium can be
written as
\begin{equation}
\hat{H}=\hat{H}_0+ez\mathscr{E}(x,t).
\end{equation}
Certainly, $x=\sqrt{\hbar/2m_e\nu}(\hat{a}+\hat{a}^\dagger)$, and
thus the above Hamiltonian can be further written as
\begin{equation}
\begin{array}{l}
\hat{H}=\hat{H}_0
+\hbar\widetilde{\Omega}\hat{\sigma}_z(e^{i\eta(\hat{a}+\hat{a}^\dagger)-i\omega_lt+i\phi_l}
+e^{-i\eta(\hat{a}+\hat{a}^\dagger)+i\omega_lt-i\phi_l})\\
\,\,\,\,\,\,\,\,\,\,\,\,\,+\hbar\Omega(\hat{\sigma}_-+\hat{\sigma}_+)
(e^{i\eta(\hat{a}+\hat{a}^\dagger)-i\omega_lt+i\phi_l}+e^{-i\eta(\hat{a}+\hat{a}^\dagger)+i\omega_lt-i\phi_l}),
\end{array}
\end{equation}
with $\Omega=\langle g|z|e\rangle eE_z/(2\hbar)$ being the so-called
carrier Rabi frequency describing the strength of coupling between
the applied laser field and the electron, and
$\widetilde{\Omega}=(\langle e|z|e\rangle-\langle
g|z|g\rangle)eE_z/(4\hbar)\neq 0$ due to the broken parities of the
quantum states of the above 1D hydrogen. Also,
$\eta=k_l\sqrt{\hbar/2m_e\nu}$ is the so-called LD parameter, which
describes the strength of coupling between the motions of $z$- and
$x$-directions of the trapped electron.
Finally, $\hat{\sigma}_-=|g\rangle\langle e|$ and
$\hat{\sigma}_+=|e\rangle\langle g|$ are the usual raising and
lowering operators, respectively. In the interaction picture defined
by $\hat{U}(t)=\exp[(-i/\hbar)\hat{H}_0t]$, the Hamiltonian (7)
reduces to
\begin{equation}
\begin{array}{l}
\hat{H}_{\text{I}}=\hbar\widetilde{\Omega}e^{i\phi_l}\hat{\sigma}_ze^{-i\omega_lt}e^{i\eta(\hat{a}e^{-i\nu
t}+\hat{a}^\dagger e^{i\nu t})}
+\hbar\widetilde{\Omega}e^{-i\phi_l}\hat{\sigma}_ze^{i\omega_lt}e^{-i\eta(\hat{a}e^{-i\nu
t}+\hat{a}^\dagger e^{i\nu
t})}\\
\,\,\,\,\,\,\,\,\,\,\,\,\,\,+\hbar\Omega
e^{i\phi_l}\hat{\sigma}_-e^{-it(\omega_0+\omega_l)}e^{i\eta(\hat{a}e^{-i\nu
t}+\hat{a}^\dagger e^{i\nu t})}+\hbar\Omega
e^{-i\phi_l}\hat{\sigma}_-e^{-it(\omega_0-\omega_l)}e^{-i\eta(\hat{a}e^{-i\nu
t}+\hat{a}^\dagger e^{i\nu t})}\\
\,\,\,\,\,\,\,\,\,\,\,\,\,\,+\hbar\Omega
e^{i\phi_l}\hat{\sigma}_+e^{it(\omega_0-\omega_l)}e^{i\eta(\hat{a}e^{-i\nu
t}+\hat{a}^\dagger e^{i\nu t})}+\hbar\Omega
e^{-i\phi_l}\hat{\sigma}_+e^{it(\omega_0+\omega_l)}e^{-i\eta(\hat{a}e^{-i\nu
t}+\hat{a}^\dagger e^{i\nu t})}.
\end{array}
\end{equation}

Now, we assume that the frequencies of the applied laser fields are
sequentially set as $\omega_l=\omega_0+K\nu$ with $K=0,\pm1$
corresponding to the usual resonance ($K=0$), the first blue-($K=1$)
and red-($K=-1$) sidebands excitations~\cite{Weistates},
respectively.
The LD parameters introduced above become
$\eta=(\omega_0+K\nu)\sqrt{\hbar/(2m_e\nu)}/c$ (where $c$ is the
velocity of light) and are sensitive to the frequencies $\omega_0$
and $\nu$, which are further relative to the applied trap field
$E_\perp$ and the depth $h$ of the micro-electrode set beneath the
liquid Helium surface.
Under the well-known LD approximation~\cite{Rev.Mod.trapped.ion}
with $\eta\ll1$, we have $\exp[\pm i\eta(\hat{a}e^{-i\nu
t}+\hat{a}^\dagger e^{i\nu t})]\approx1\pm i\eta(\hat{a}e^{-i\nu
t}+\hat{a}^\dagger e^{i\nu t})$ and simplify the above Hamiltonian
to
\begin{equation}
\begin{array}{l}
\hat{H}_{\text{I}}^{\text{L}}(t)=
\hbar\widetilde{\Omega}e^{i\phi_l}\hat{\sigma}_z[e^{-i(\omega_0+K\nu)t}+i\eta(\hat{a}e^{-i(\omega_0+K\nu+\nu)
t}+\hat{a}^\dagger e^{-i(\omega_0+K\nu-\nu) t})]
\\

\,\,\,\,\,\,\,\,\,\,\,\,\,\,\,\,\,\,\,\,\,

+\hbar\widetilde{\Omega}e^{-i\phi_l}\hat{\sigma}_z[e^{i(\omega_0+K\nu)t}-i\eta(\hat{a}e^{i(\omega_0+K\nu-\nu)
t}+\hat{a}^\dagger e^{i(\omega_0+K\nu+\nu) t})]\\
\,\,\,\,\,\,\,\,\,\,\,\,\,\,\,\,\,\,\,\,\,

+\hbar\Omega e^{i\phi_l}\hat{\sigma}_-
[e^{-i(2\omega_0+K\nu)t}+i\eta(\hat{a}e^{-i(2\omega_0+K\nu+\nu)
t}+\hat{a}^\dagger e^{-i(2\omega_0+K\nu-\nu) t})]
\\

\,\,\,\,\,\,\,\,\,\,\,\,\,\,\,\,\,\,\,\,\,

+\hbar\Omega e^{-i\phi_l}\hat{\sigma}_-[e^{iK\nu
t}-i\eta(\hat{a}e^{i(K-1)\nu
t}+\hat{a}^\dagger e^{i(K+1)\nu t})]\\

\,\,\,\,\,\,\,\,\,\,\,\,\,\,\,\,\,\,\,\,\,

+\hbar\Omega e^{i\phi_l}\hat{\sigma}_+ [e^{-iK\nu
t}+i\eta(\hat{a}e^{-i(K+1)\nu
t}+\hat{a}^\dagger e^{i(1-K)\nu t})]\\

\,\,\,\,\,\,\,\,\,\,\,\,\,\,\,\,\,\,\,\,\,

+ \hbar\Omega
e^{-i\phi_l}\hat{\sigma}_+[e^{i(2\omega_0+K\nu)t}-i\eta(\hat{a}e^{i(2\omega_0+K\nu-\nu)
t}+\hat{a}^\dagger e^{i(2\omega_0+K\nu+\nu) t})].

\end{array}
\end{equation}
Neglecting the above rapidly-oscillating terms (i.e., under the
usual rotating-wave approximation)~\cite{RWA}, this Hamiltonian can
be further simplified to
\begin{equation}
\hat{H}_{\text{eff}}^{0}=\hbar\Omega
e^{i\phi_l}\hat{\sigma}_++H.c\,\,\,\,\,\,\,\,\,\,\,\,\,\,\,\,\,\,\,\text{for}\,\,\,K=0,
\end{equation}
\begin{equation}
\hat{H}_{\text{eff}}^{\text{r}}=i\eta\hbar\Omega
e^{i\phi_l}\hat{\sigma}_+\hat{a}+H.c\,\,\,\,\,\,\,\,\,\,\text{for}\,\,\,K=-1,
\end{equation}
\begin{equation}
\hat{H}_{\text{eff}}^{\text{b}}=i\eta\hbar\Omega
e^{i\phi_l}\hat{\sigma}_+\hat{a}^\dagger+H.c\,\,\,\,\,\,\,\,\,\,\text{for}\,\,\,K=1.
\end{equation}

Obviously, Hamiltonians $\hat{H}_{\text{eff}}^{\text{r}}$ and
$\hat{H}_{\text{eff}}^{\text{b}}$ are nothing but just those of the
usual JCM and anti-JCM, respectively.
All the dynamical evolutions corresponding to the above effective
Hamiltonians (10-12) are exactly solvable.
For example, if the $x$-direction's harmonic oscillator is prepared
initially at the Fock state $|m\rangle$ ($m$ is its occupation
number), then we have

i) For $K\leq0$
\begin{eqnarray}
\left\{
\begin{array}{l}
|m\rangle|g\rangle\longrightarrow|m\rangle|g\rangle,\,\,m<k,\\
|m\rangle|g\rangle\longrightarrow\cos(\Omega_{m-k,k}t)|m\rangle|g\rangle
+i^{k-1}e^{i\theta_{L}}\sin(\Omega_{m-k,k}t)|m-k\rangle|e\rangle;\,\,m\geq
k,\\
|m\rangle|e\rangle\longrightarrow\cos(\Omega_{m,k}t)|m\rangle|e\rangle
-(-i)^{k-1}e^{-i\theta_{L}}\sin(\Omega_{m,k}t)|m+k\rangle|g\rangle
\end{array}
\right.
\end{eqnarray}

ii) For $K\geq0$
\begin{eqnarray}
\left\{
\begin{array}{l}
|m\rangle|g\rangle\longrightarrow\cos(\Omega_{m,k}t)|m\rangle|g\rangle
+i^{k-1}e^{i\theta_{L}}\sin(\Omega_{m,k}t)|m+k\rangle|e\rangle,\\

|m\rangle|e\rangle\longrightarrow|m\rangle|e\rangle,\,m<k,\\

|m\rangle|e\rangle\longrightarrow\cos(\Omega_{m-k,k}t)|m\rangle|e\rangle
-(-i)^{k-1}e^{-i\theta_{L}}\sin(\Omega_{m-k,k}t)|m-k\rangle|g\rangle,\,m\geq
k,
\end{array}
\right.
\end{eqnarray}
with $\Omega_{m,k}=\Omega\eta^{k}\sqrt{(m+k)!/m!}$ being the
effective Rabi frequency, and $k=|K|$.
In principle, arbitrary quantum state engineering, e.g., generations
of nonclassical quantum states and implementations of quantum logic
gates, etc.~\cite{gates,JCMs1,Rev.Mod.trapped.ion}, could be
realized by the above evolutions.

The experimental feasibility of the JCMs proposed here involves with
two important factors: the value of the introduced LD parameter
$\eta$ and the decoherence of the electron. In fact, decoherence is
always a challenge in various quantum coherence systems. Platzman
and Dykman~\cite{PRB} showed that the main source of decoherence in
the present system is the so-called ripplons, i.e, the thermally
excited surface waves of liquid Helium~\cite{PRB,Science}. The
coherence time due to this fluctuation is
estimated~\cite{PRB,Science} to be $10^{-4}$~s (for the typical
frequencies: a few tens of GHz), but could be increased by enhancing
the frequency of the electron vibrating in-plane.

For the typical parameters $E_\perp\approx10^{4}$V/m and
$h\approx5\times10^{-7}$m~\cite{PRB}, the transition frequency of
the $z$-direction's 1D hydrogen and the vibrational frequency of the
$x$-direction's oscillation are estimated as
$\omega_0\approx1133$~GHz and $\nu\approx59$~GHz, respectively.
Consequently, the LD parameter in the above JCM is
$\eta\approx1.2\times10^{-4}\ll 1$. Thus, the usual LD approximation
is valid. Note that the LD parameter in present system is
significantly smaller than that (there $\eta\sim 0.2$) in the
experimental ion trap system~\cite{JCMs1,Rev.Mod.trapped.ion}.
This is because the ``atomic" frequency $\omega_0$ of the trapped
ion ($\sim 10^{6}$ GHz) is significantly {\it larger} than that in
the present system ($\sim 1 $THz), and the vibrational frequency
$\nu$ ($\sim 10^{-4}$ GHz) is significantly however {\it smaller}
than that in the present system.
Note that the LD parameters could, in principle, been enlarged by
decreasing the value of $\nu$ (by properly adjusting $E_\perp$ and
$h$, e.g., $\eta\approx0.16$, $\omega_0\approx739$~GHz,
$\nu\approx13.3$~KHz for $E_\perp=10^{-5}$~V/m and $h=10^{-2}$~m).
However, ripplons-induced decoherence affects stronger for the
smaller frequency of the vibrations in the plane (correspond to a
large in-plane localization length).

Fortunately, although the LD parameters in present system are
relatively small, the JCMs presented above still work within the
typical coherence time ($\sim 10^{-4}$s). Our numerical estimations
show that the duration of a $\pi$-pulse is
$t=\pi/\Omega_{m,k}<10^{-4}$s. For example, if the amplitude of the
applied laser field is set as the typical value:
$E_z=10^2$V/m~\cite{PRB}, and the LD parameter
$\eta=1.2\times10^{-4}$, we have
$\pi/\Omega_{m,0}\approx9.1\times10^{-9}$s and
$\pi/\Omega_{m,1}<7.6\times10^{-5}$s. Note that the occupation
number $m$ does not affect the values of $t=\pi/\Omega_{m,0}$, while
$t=\pi/\Omega_{m,1}$ decreases with the increase of $m$
($t\propto1/\sqrt{m+1}$).
Also, the above durations could be further shortened (such that the
JCMs admit more $\pi$-pulse operations) by effectively increasing
the amplitude $E_z$ of the applied laser field (i.e., increasing the
carrier Rabi frequency $\Omega$). In principle, if the $E_z$
increases ten times, then the duration of a $\pi$-pulse shortens ten
times. Indeed, for $E_z=10^3~V/m$ a $\pi$-pulse could be less than
$7.6\times10^{-6}$~s.

The standard JCM requires that the bosonic field should be in a pure
state. However, thermal states
\begin{equation}
\rho_{t}=\sum_{m=0}^\infty \frac{\langle m\rangle^m}{(1+\langle
m\rangle)^{m+1}}|m\rangle\langle m|,\,\,\,\,\langle
m\rangle=\frac{1}{e^{\hbar\nu/(k_BT)}-1},
\end{equation}
are the natural states of the vibrational particles (e.g., trapped
ions~\cite{JCMs1} and the electrons in the present model), which are
normally in thermal equilibrium with their surroundings.
\begin{figure}[tbp]
\includegraphics[width=9.5cm]{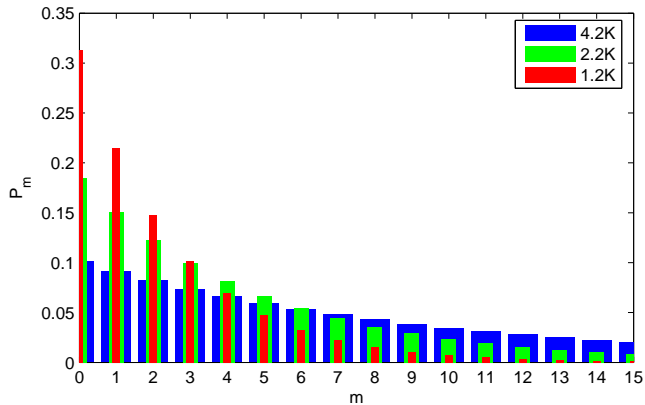}
\caption{(Color online) Phonon distributions $P_m$ versus occupation
number $m$ of (vibrational) frequency $\nu=59$GHz in the thermal
states for the typical temperatures $T=4.2$K,\,$2.2$K, and $1.2$K.}
\end{figure}
Above, $k_B$ and $T$ are the Boltzmann constant and the temperature
of the surroundings, respectively. Fig.2 shows the phonon
distribution of a thermal state for a vibration with frequency
$\nu=59$\,GHz (corresponding to $0.45$K) at various typical
temperatures: $T=4.2$K,\,$2.2$K, and $1.2$K. Obviously, if the
temperature of the surrounding is further lower, the probabilities
that the electron in the states with smaller occupations are much
larger.
Suppose that the liquid Helium is cooled to $T=0.01$K~\cite{PRB},
which is much colder than $0.45$K of the vibrational
electron~\cite{PRB}, then $\rho_{t}\approx|0\rangle\langle0|$ and
thus the electronic vibration is well limited to the vacuum state.

In addition, the presented JCMs could be utilized to cool the
vibrational electron. Indeed, if the out-plane state of the electron
is initially in $|g\rangle$, a vibrational energy $\hbar\nu$ could
be reduced by the following two steps: (i) apply a $\pi/2$-pulse
with duration $t=\pi/2\Omega_{m-1,1}$ to drive a transition:
$|m\rangle|g\rangle \rightarrow|m-1\rangle|e\rangle$; (ii) drive the
transition $|e\rangle \rightarrow |g\rangle$ but forbid the
transition $|g\rangle \rightarrow |e\rangle$ by using an auxiliary
atomic level $|a\rangle$ and two resonant $\pi/2$-pulses to drive
the transitions $|e\rangle\rightarrow|a\rangle\rightarrow|g\rangle$.
For example, if the third level of the electron is selected to be
the auxiliary level $|a\rangle$, we have
$\omega_{ea}/\omega_{ge}\approx 0.18$ with $\omega_{ea}$ being the
transition frequency between $|e\rangle$ and $|a\rangle$ and
$\omega_{ge}$ between $|g\rangle$ and $|e\rangle$. After these two
steps, cooling the vibrational electron by a $\hbar\nu$ is possible:
\begin{equation}
|m\rangle|g\rangle{\longrightarrow}
|m-1\rangle|e\rangle{\longrightarrow}|m-1\rangle|a\rangle
{\longrightarrow}|m-1\rangle|g\rangle.
\end{equation}
These operations (their durations are typically less than
$4\times10^{-6}$~s for $E_z=10^3$V/m) are repeated until the
vibrational state $|m\rangle$ relaxes finally to the desirable
ground state $|0\rangle$. As a consequence, an arbitrary mix state
$\rho=\sum_mP_m|m\rangle\langle m|$ (with $P_m$ being the classical
probability that the electron is in the vibrational state
$|m\rangle$) could be cooled to the vibrational vacuum state
$|0\rangle$. Note that the above method is similar to the so-called
{\it sideband laser cooling} technique used usually in the trapped
ion system~\cite{Rev.Mod.trapped.ion}.

In conclusion, we have proposed a new candidate to realize the
famous JCMs: {\it electrons on liquid Helium}, by applying classical
laser fields to the trapped electrons for coupling their motions
along the $x$- and $z$-directions. We have shown that the desirable
JCMs and anti-JCMs could be implemented by properly setting the
frequencies of the applied laser beams to excite the first red- and
blue sidebands, respectively. The present proposal provides a new
way to apply the famous JCMs in condensed matters.

{\bf Acknowledgements}: This work is partly supported by the NSFC
grant No. 10874142 and the grant from the Major State Basic Research
Development Program of China (973 Program) (No. 2010CB923104).


\end{document}